%% file: Jan5AB.tex
\RequirePackage{lineno}

\documentclass[prd,twocolumn,preprintnumbers,superscriptaddress,nofootinbib]{revtex4}

\usepackage[title]{appendix}
\usepackage{graphicx}  
\usepackage{dcolumn}   
\usepackage{bm}        
\usepackage{amssymb}   
\usepackage{multirow}
\usepackage{lineno}
\usepackage{pdfpages}
\usepackage[hyperfootnotes=false]{hyperref}
\usepackage{capt-of}
\usepackage{float}
\usepackage{subcaption}
\usepackage[fleqn]{amsmath}

\begin{document}

\newcommand{\carbon}{\rm ^{12}C}
\newcommand{\deuterium}{\rm ^{2}H}
%

\title{Investigation of Medium Modifications to $^{12}$C Structure Functions in the Resonance Region}
\input{authors}

\date{\today}

\begin{abstract}
We present results from a high precision experimental study of the nuclear modification of the longitudinal ($F_L$) to transverse ($F_1$) structure function ratio for bound nucleons in the resonance region.  The inclusive electron scattering cross sections were measured in Jefferson Lab Experimental Hall C on carbon and deuterium nuclei for a large range of kinematics, allowing for separations of the longitudinal and transverse structure functions to be  performed at a range of four-momentum transfer values $0.5 \le Q^2 \le$ 3.75~GeV$^2$.  In contrast to the significant body of measurements of the nuclear modification of the $F_2$ structure function in the deep inelastic scattering region, there is very little on $F_L$ and $R = F_L / 2xF_1$ in the region of the nucleon resonances.  In this Letter we present measurements of the nuclear effect on $R$ for $^{12}$C ($R_C$) relative to deuterium ($R_D$). These results indicate regions in which in $R_C>R_D$, requiring that the nuclear modifications be different in all three structure functions,  $F_2$, $F_1$ and $F_L$.

 \end{abstract}
\maketitle 
\clearpage
\newpage
\pagebreak

In 1983,  muon~\cite{EuropeanMuon:1983bvx} and electron~\cite{Bodek:1983qn} deep-inelastic scattering (DIS) experiments revealed that the quark distributions in the nucleon are modified when the nucleon is bound in a nucleus.  Numerous theoretical models were proposed to explain these modifications (see reviews~\cite{Barone:1992ej,Geesaman:1995yd,Norton:2003cb,Malace:2014uea}), commonly referred to as the {\it EMC effect}.  In the ensuing years, many measurements of the ratio of electron scattering cross sections on nuclear targets to those on deuterium ($\sigma^A/\sigma^D$) were performed \cite{Bodek:1983ec,Gomez:1993ri,Seely:2009gt,Arrington:2021vuu} as a function of the Bjorken scaling variable $x$, the square of the four-momentum transfer ($Q^2$) and the mass (W) of the hadronic final state.  These data were carried out predominately at kinematics in which the cross section is dominated by the $F_2$ structure function, precluding an examination of whether the modifications are the same in both the longitudinal ($F_L$) and transverse ($F_1$) structure functions.  The only experiment that measured the nuclear dependence of $R=F_L/2xF_1$ was  SLAC E140~\cite{Dasu_PRD94}, which did not apply Coulomb corrections and did not include a carbon target in their measurements. Those measurements were performed in the DIS region dominated by scattering from quarks, and the analysis extracted $\Delta R = R_A - R_D$ from Fe and Au nuclei.  While a nonzero $\Delta R$ would indicate a difference in $F_2^A/F_2^D$, $F_1^A/F_1^D$ and $F_L^A/F_L^D$, the analysis indicated that $\Delta R= 0$ within the systematic uncertainties. While existing world data on $R$ are largely confined in the DIS region, measurements in the nucleon resonance region are essential for understanding its nuclear dependence, and hence that of $F_L$.

The differential cross section for scattering an unpolarized charged lepton with energy  $E_0$, final energy $E^{\prime}$ and scattering angle $\theta$ can be written in terms of the structure functions $F_1$ and $F_2$ as:
\begin{flalign}
& \frac{d^2\sigma}{d\Omega\, dE'}(E_0,E',\theta)
= \frac{4\alpha^2 E'^2}{Q^4}\cos^2(\theta/2) \nonumber \\
& \quad \times \left[F_2(x,Q^2)/\nu
+ 2\tan^2(\theta/2)F_1(x,Q^2)/M\right] &&
\end{flalign}
where $\alpha$ is the fine structure constant, $M$ is the nucleon mass, $\nu=E_0-E^{\prime}$, and  $Q^2=4E_0E^{\prime} \sin ^2 (\theta/2)$. 

In quantum chromodynamics (QCD), $F_2(x,Q^2)$ in the deep-inelastic region is expressed in terms of charge weighted sums of the  fractional momentum distributions of quarks and antiquarks in the nucleon.
Within the quark parton model 
$ x=Q^2/2M\nu$ is the fractional momentum (parallel to the direction of the momentum transfer) carried by the struck quark in the nucleon. 

Alternatively, one can view this scattering process in terms of the cross section for the absorption of transverse $(\sigma_T)$ and longitudinal $(\sigma_L)$ virtual photons, where 
\begin{flalign}
\frac{d^2\sigma}{d\Omega dE^\prime} &=
  \Gamma \left[\sigma_T(x,Q^2) + \epsilon \sigma_L(x,Q^2) \right] \\
       & =  {4 \pi^2 \alpha \over K M} \times \Gamma \left[F_1(x,Q^2) + \epsilon {F_L(x,Q^2) \over 2x} \right]
     \label{req}   
     \end{flalign}
 Here, $ \Gamma = \frac{\alpha K E^\prime}{ 4 \pi^2 Q^2 E_0}  \left( \frac{2}{1-\epsilon } \right)$ is the virtual photon flux,  $K = \frac{2M \nu - Q^2 }{2M}$, and  $\epsilon = \left[ 1+2(1+\frac{Q^2}{4 M^2 x^2} ) \tan^2{ \frac{\theta}{2}} \right] ^{-1}$ is the relative flux of longitudinal virtual photons in the Hand convention~\cite{hand-PhysRev129}.
The structure functions $F_1$, $F_L$, and $F_2$ are proportional to $\sigma_T$, $\sigma_L$, and $[\sigma_T + \sigma_L]$, respectively. The ratio    
$ R=\sigma_L/\sigma_T$ is related to the structure functions by,
\begin{equation}\label{eq:R_eq}
 R(x,Q^2)
   \equiv \frac {\sigma_L }{ \sigma_T}
   = \frac{F_2 }{ 2x F_1}(1+\frac{4M^2x^2 }{Q^2})-1
   = \frac{F_L }{ 2x F_1}\; .
\end{equation}

Contributions to $R$ originate from the  perpendicular component of  the momentum of the spin 1/2 quarks with respect to the momentum transfer vector \cite{PhysRevD.5.528,PhysRevD.20.1471,PhysRevD.19.796,PhysRevD.27.285}.  At small $x$, $R$ is dominated by a perpendicular momentum component that originated from QCD gluon emission \cite{Martin:2000gq} ($R_{QCD}$). At large $x$, $R$ is dominated by  quark binding in the nucleon (the so-called {\it target mass corrections} \cite{Barbieri:1976rd,Georgi:1976ve}  $R_{TM}$). While at small $Q^2$, $R$ is dominated by nonperturbative processes, such as interactions with more than one quark (higher twist). In addition, at small $Q^2$  an enhancement of the longitudinal cross section can also originate from scattering from integer spin particles,  e.g. meson clouds in the nucleus \cite{Miller:2001yf,Zaidi:2017tpr}.  At very small $Q^2$,  the quark-parton model breaks down and from current conservation $R$ must be zero at $Q^2=0$.

\begin{figure*}[ht]
\begin{subfigure}[t]{0.49\textwidth}
  \centering
  \includegraphics[width=\linewidth]{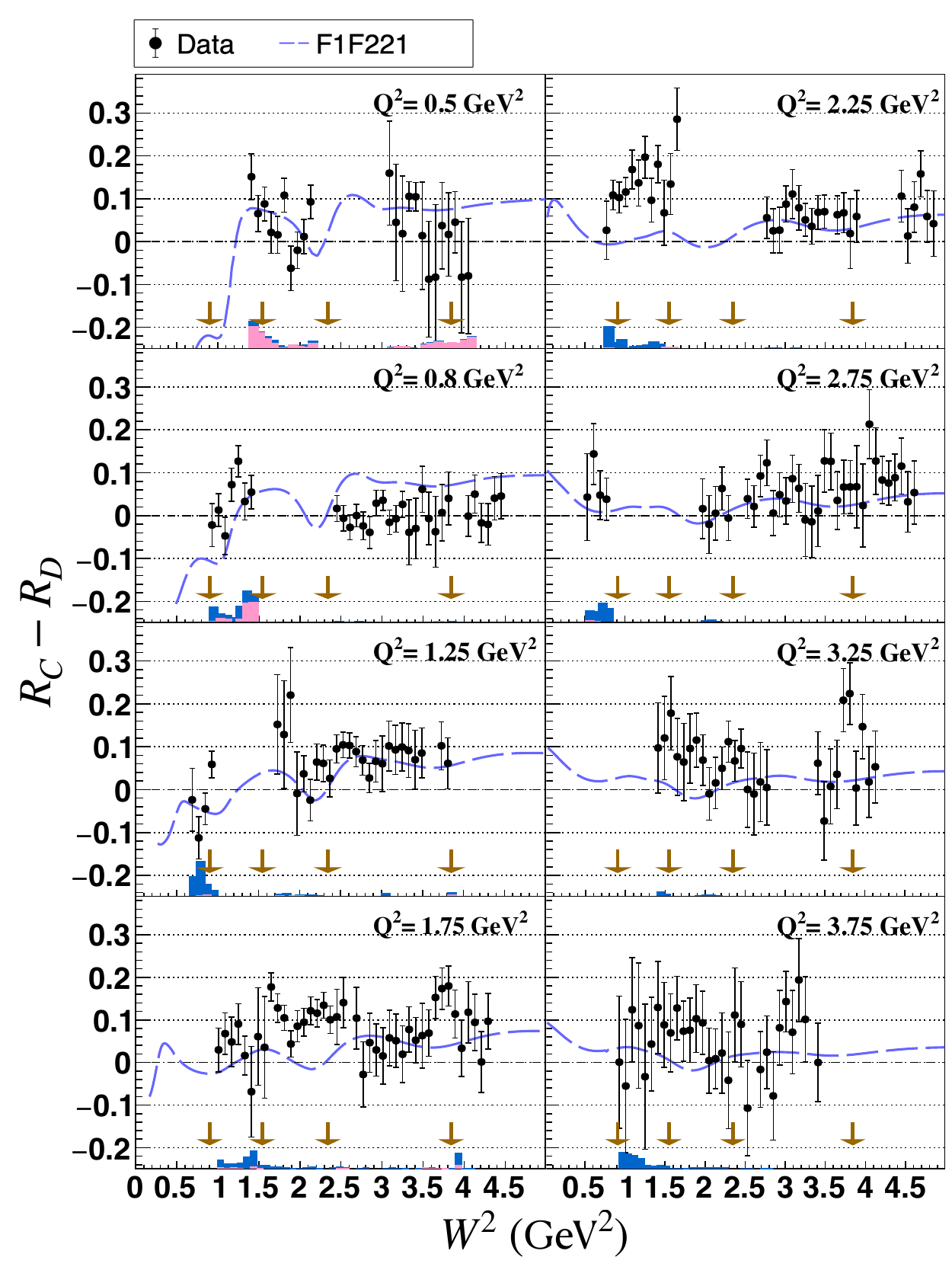}
  \caption{}
  \label{fig:dR_a}
\end{subfigure}%
\hfill
\begin{subfigure}[t]{0.49\textwidth}
  \centering
  \includegraphics[width=\linewidth]{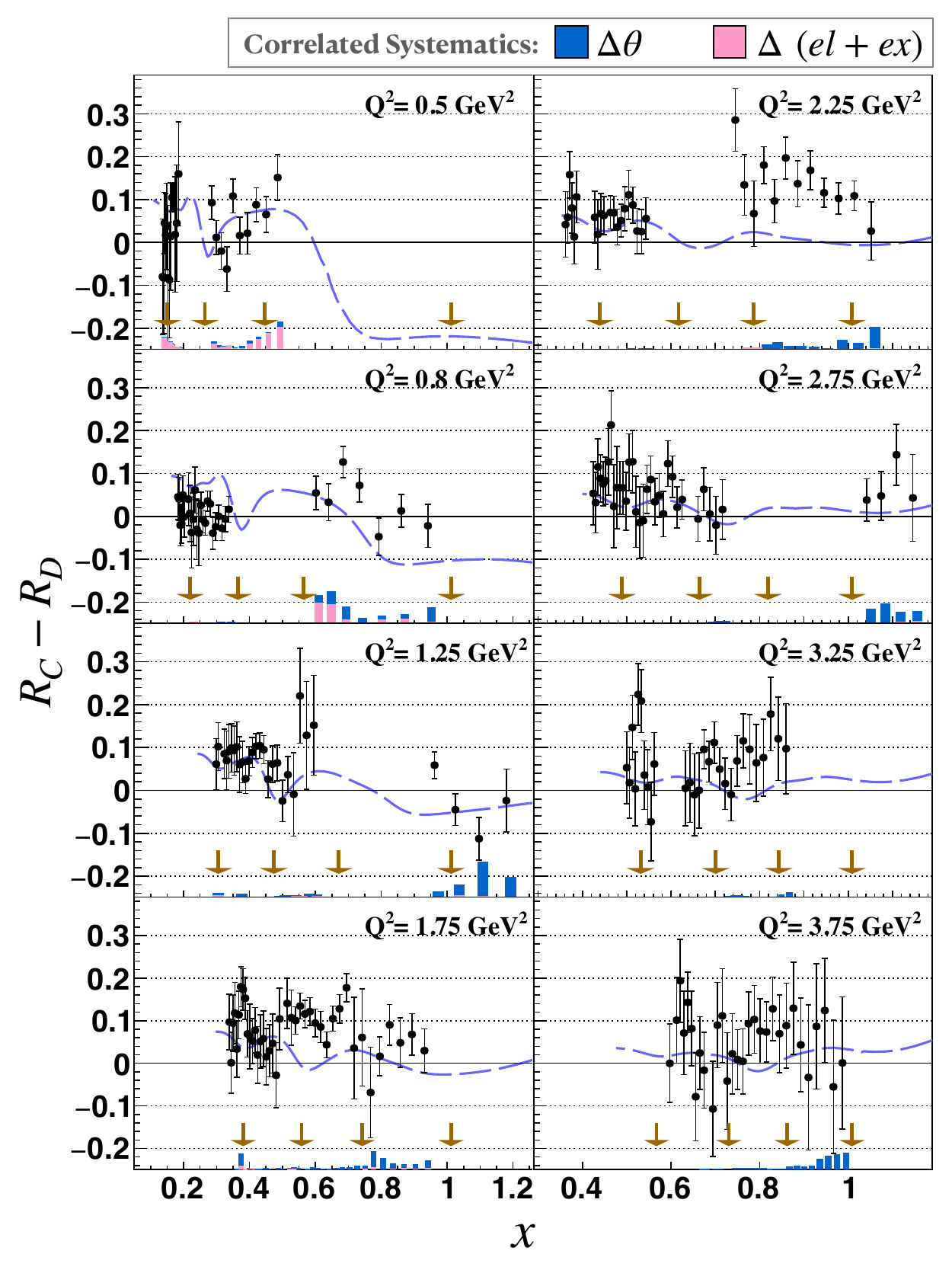}
  \caption{}
    \label{fig:dR_b}
\end{subfigure}
\caption{The extracted values of $R_C - R_D $  versus $W^2$ (a) and versus $x$ (b) for bins in $Q^2$  ($0.5 \le Q^2 \le 3.75$ GeV$^2$).   The locations \cite{Stein:1975yy} of the QE peak and the 1232, 1520 and 1950 MeV nucleon resonances are indicated by vertical arrows. The curves represent the fit \cite{Bodek:2022gli,Bodek:2023dsr} to all available experimental data on deuterium and carbon (including our iterated data), updating the parametrizations described in~\cite{Bosted:2012qc,Bosted:1994tm,Christy:2007ve,Bosted:2007xd}. The blue and magenta bars represent the correlated uncertainties from the angle offset and from the radiative tail from the excited states in $^{12}$C, respectively.}
\label{fig:Fig_DR}
\end{figure*}

 In this Letter we report on precise measurements of the $W^2$ and $Q^2$ dependence of $R = \sigma_L/\sigma_T$ for deuterium and $^{12}$C using the Hall C spectrometers at Jefferson lab, covering the same kinematic region as experiment E94-110 (which previously measured $R$  for the proton with the same apparatus~\cite{JeffersonLabHallCE94-110:2004nsn}). The data cover a range of $Q^2$ between 0.4 and 4.0 GeV$^2$ and energy  transfer ($\nu$) corresponding to $W^2=M^2+2M\nu-Q^2 <$  5.0 GeV$^2$, covering the quasielastic,  resonance  and inelastic continuum regions.  Targets include deuterium, carbon, aluminum, iron,  and copper nuclei. Here,  we focus on deuterium and carbon in the resonance region (heavier nuclei are under analysis). The dataset consists of measurement taken in 2005 and 2007 with D and $^{12}$C targets. The applied relative normalization factor of the 2007 dataset relative to that from 2005 was determined from a global fit~\cite{Bodek:2022gli,Bodek:2023dsr}  to be 1.01, which accounts for the small difference between the two datasets and remains well within the 1.7\% normalization uncertainty assigned to each.

Incident electrons at ten different energies (1.2, 2.1, 2.3, 3.12, 3.27, 3.4, 4.07, 4.13, 4.2 and 5.15 GeV) provided by the Continuous Electron Beam Accelerator Facility (CEBAF) at Jefferson Lab are scattered from a 4-cm-long liquid deuterium  target, and a $\sim$ 2\% radiation length carbon target.  Electrons are detected by the Hall C High Momentum Spectrometer (HMS) at angle settings ranging from 10.65$^\circ$ to 75$^\circ$. The experiment comprised a total of 44 angular settings and 176 momentum settings, with beam currents ranging from 30 to 80 $\mu$A. 

The charge symmetric (CS) backgrounds originating from symmetric pairs of $e^+$ and $ e^-$ produced by the conversion of photons from  $\pi^0$ production and subsequent decay are measured by reversing the HMS magnet polarities to determine the yield of $e^+$.  Background from electroproduced charged pions is identified and removed by using both a gas Cherenkov counter and an electromagnetic calorimeter.  Events originating from electron scattering off the aluminum walls of the cryogenic target cell are subtracted by measuring the yield from an empty target replica~\cite{BODEK1973603}. For additional details regarding this analysis see~\cite{Alsalmi2019,Albayrak2011,Mamyan:2012th}. For details on measurements of electron scattering cross sections with the Hall C apparatus see  \cite{E94110:2004lsx}. \par
The differential cross sections are determined from the  background-corrected electron yields after correcting for inefficiencies and radiative corrections.  Radiative effects are evaluated with the exact Mo-Tsai formalism~\cite{RevModPhys.41.205}, modified to include quark loop corrections~\cite{Whitlow:1990dr,Whitlow:1990gk}. They also include bremsstrahlung, vertex corrections and loop diagrams standard to electron scattering experiments. In all previous electron scattering experiments on nuclear targets at high energies (at SLAC and Jefferson Lab) the contribution from the radiative tail of nuclear excitations was neglected.  In this analysis we correct for these using a fit to the  $^{12}$C nuclear excitation form factors~\cite{Bodek:2023dsr} because we find that it cannot be neglected at low $Q^2$ and large $\nu$. 

In the calculation of radiative corrections (and Coulomb corrections described below) we model the electron scattering cross section by a universal fit to all available electron scattering data (including our data) on hydrogen, deuterium and carbon~\cite{Bodek:2022gli,Bodek:2023dsr}.  Since the fit is used to determine the radiative, Coulomb, and bin-center corrections, we iterate the extraction and fit three times, achieving convergence less than 0.2\% between iterations. 

\begin{figure*}[ht]
\begin{subfigure}[t]{0.49\textwidth}
  \centering
  \includegraphics[width=\linewidth]{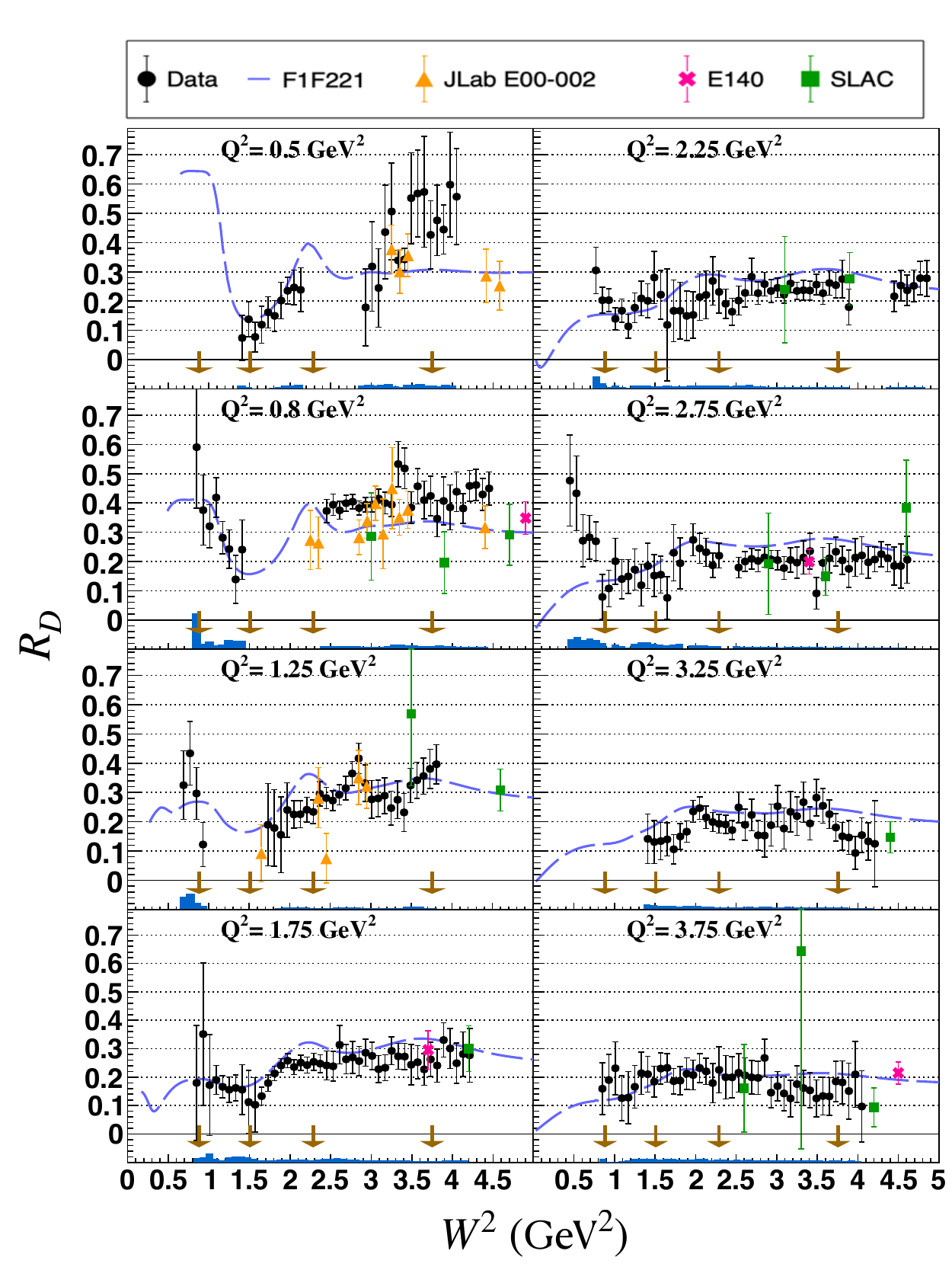}
  \caption{}
  \label{fig:R_d}
\end{subfigure}%
\hfill
\begin{subfigure}[t]{0.49\textwidth}
  \centering
  \includegraphics[width=\linewidth]{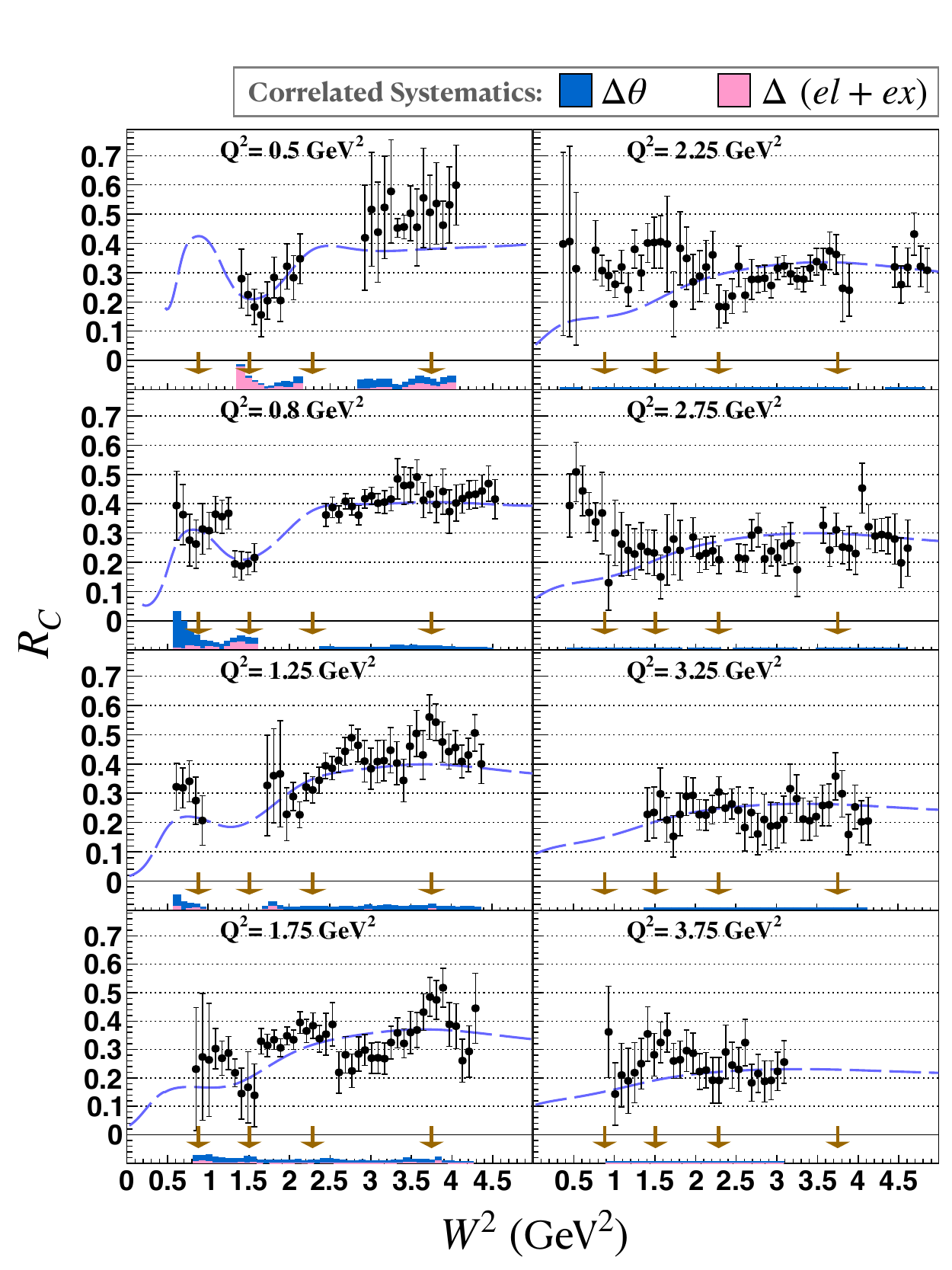}
  \caption{}
    \label{fig:R_c}
\end{subfigure}
\caption{As in Fig.~\ref{fig:Fig_DR}, but showing $R_D$ (a) and $R_C$ (b) separately versus $W^2$ in the same $Q^2$ bins.}
\label{Fig_R}
\end{figure*}

For targets with atomic number $Z>1$, Coulomb corrections are applied to account for the effect of the  nuclear electric field on the incident and scattered electrons. Within  the effective momentum approximation (EMA)~\cite{Aste:2005wc}, this is implemented using an  effective potential $V_{eff}$, defined as: $V_{eff}=0.75-0.8 V_0$, where $V_0$ is the central electrostatic potential~\cite{Solvignon_2009}.  This prescription is consistent with values extracted from comparisons of positron and electron scattering cross sections; for example: for $^{12}$C one finds $V_{eff}$=3.1$\pm$0.25~MeV~\cite{Gueye:1999mm}. 
At the scattering vertex, the effective incident and scattered electron energies are shifted to $E_{eff}=E_0+V_{eff}$ and $E'_{eff}= E'+V_{eff}$, respectively.  In addition, a focusing factor $F_{foc}=(E_0+V_{eff})/E_0$ is applied. 

The model of the differential cross section  ($\sigma_{model}$) is used to correct the measured cross sections $\sigma_{meas}$ and yield a Coulomb corrected cross section ($\sigma^{cc}_{meas}$):

  \begin{equation}
\sigma^{cc}_{meas}(E_0,E') = \frac {\sigma_{meas}(E_0,E')} {F^2_{foc} }   \times
   \frac {\sigma_{model}(E_0,E') }{\sigma_{model}(E_{eff},E'_{eff})}.
\end{equation}

The measured cross section in fixed bins of $W^2$ is interpolated to fixed $Q^2$ values of 0.5, 0.8, 1.25, 1.75, 2.25, 2.75, 3.25 and 3.75 GeV$^2$ utilizing the global fit. 

To probe the nuclear dependence of the longitudinal and transverse structure functions, we form the ratio of deuterium to carbon cross sections, which can be written as
  \vspace{-5pt}
\begin{eqnarray}
\frac{\sigma_D}{\sigma_C} &=& 
\frac{\sigma_D^T+\epsilon \sigma_D^L}{\sigma_C^T+\epsilon \sigma_C^L}= \frac{\sigma_D^T}{\sigma_C^T}\frac{1 +\epsilon R_C-\epsilon R_C+\epsilon R_D}{1+\epsilon R_C}  \nonumber\\
&=&\frac{\sigma_D^T}{\sigma_C^T} \left[1 - \epsilon'(R_C-R_D) \right]\,
\label{req2}
\end{eqnarray}
where  $\epsilon'   = \epsilon /(1+\epsilon R_C)$.   Note that  $\epsilon R_C$ is small and the resonance structure in $R_C$ is smeared by Fermi motion. The values of $R_C$ used here are taken from the fit of Refs. ~\cite{Bodek:2022gli,Bodek:2023dsr}. 

  
  Fig.~\ref{fig:Fig_DR} shows  $R_C-R_D$ versus $W^2$ [Fig.~\ref{fig:dR_a}]  and versus $x$ [Fig.~\ref{fig:dR_b}] for eight different values of $Q^2$. A requirement of $\Delta \epsilon' \ge 0.2$ was imposed, where $\Delta \epsilon'$ is the difference between the largest and smallest $\epsilon'$ values used in each fit. Gaps indicate bins with $\Delta \epsilon' < 0.2$ or no data. The data clearly demonstrate a significant difference in the nuclear modifications of the longitudinal and transverse structure functions of bound nucleons in the resonance region, with $R_C$ on average exceeding $R_D$ by $\approx$0.062 (or $\approx$25\%). \par

 This observation is in contrast with expectations at moderate and large $x$. For $x>0.3$ the contribution to $R$ are primarily from target mass corrections and the gluon contribution is small. Therefore, since nucleon target mass corrections are not changed by nuclear effects  it is expected \cite{Armesto:2010tg} that  $R_A=R_D$ for most of the $x$ region of our measurements. Theoretical calculations of Fermi-motion effects on the longitudinal structure function~\cite{Ericson:2002ep} also predict a very small difference with $R_C$ exceeds $R_D$ by only 5\% for $x>0.4$.  \par
 
Three types of systematic uncertainties are considered: point-to-point, correlated and normalization. Point-to-point uncertainties can vary with $x$ and $Q^2$, while normalization uncertainties affect only the overall scale. 
Most systematic effects, including normalization uncertainties,  cancel in the linear fit extraction of $R_C-R_D$ (Eq.~\ref{req2})~\cite{PhysRevD.20.1471}.
However, the cross sections entering Eq.~(\ref{req2}) retain statistical and uncorrelated point-to-point uncertainties in $\epsilon'$ that do not cancel. Residual $\epsilon'$-correlated systematics are dominated by (i) a spectrometer angle offset and (ii) the radiative tail from the excited states in $^{12}$C. The angle offset is constrained to 0.2~mrad shift by surveys and optics calibrations. Its impact is evaluated by re-extracting the cross sections with a 0.2~mrad shift, which produces a kinematic-dependent variation in $R_C-R_D$. 

 Radiative corrections do not cancel in the $^{12}$C to deuterium ratio due to the differences in elastic form factors and the presence of nuclear excited states in $^{12}$C.  Beam bremsstrahlung enhances contributions from these excitations at large $\nu$ and low $Q^2$.  The present analysis utilized recent fits ~\cite{Bodek:2023dsr} to the global data on nuclear elastic and excitation form factors to evaluate these effects. Based on the consistency of the fits with existing data, uncertainties of 5\% and 15\%  are assigned to the elastic and excitation radiative tail.  In addition, we tested the effect of using the Guthrie Miller formalism\cite{Zielinski:2017,Miller:1971qb,Miller:1971dj} for the elastic tail instead of the Mo-Tasai approach and find that the difference is within the quoted uncertainties. The uncertainty from external radiative correction is estimated to be negligible~\cite{Dasu:1988bd}.

\begin{table}[]
    \centering   
        \caption{The average values of $ R_C-R_D$ and $R_D$ over the region $1.5\le W^2\le4.75$ GeV$^2$ for $ 0.5 \le Q^2  \le 3.75$   GeV$^2$.}
    \begin{tabular}{|>{\centering}m{0.7in}| >{\centering}m{.95in}  |>{\centering\arraybackslash}m{0.85in}|}
     \hline
  $Q^2$ (GeV$^2$)   &  $\langle R_C-R_D\rangle$ &  $\langle R_D \rangle$ \\
  \hline
     0.5 &  $0.053 \pm  0.012$  & $0.266\pm 0.014$    \\
      0.8 &  $0.005\pm 0.008$   & $0.402 \pm 0.008$   \\
     1.25 & $0.073 \pm 0.009$  &  $0.289\pm0.009$     \\
     1.75 &  $0.101 \pm 0.008$  & $0.240 \pm 0.007$    \\
     2.25 &   $0.070\pm 0.012$    & $0.232\pm 0.008$    \\
    2.75 &   $0.058 \pm 0.011$ & $0.202 \pm 0.008$      \\
    3.25 &  $0.069 \pm 0.013$  &  $0.190 \pm 0.008$   \\
    3.75 &   $0.059 \pm 0.018$  & $0.188 \pm 0.010$  \\
       \hline
    \end{tabular}
    \label{tab:ave}
\end{table}

The average values of  $\Delta R = R_C-R_D$ and $R_D$ over the region $1.5\le W^2\le4.75$ GeV$^2$ are shown in Table \ref{tab:ave}, for $Q^2$ values of  0.5, 0.8,  1.25, 1.75, 2.25,  2.75,  3.25, and 3.75  GeV$^2$,  respectively. The values and uncertainties of the points and error bars shown in the figures are included in Supplementary Material \cite{Supplemental}.\par 
\begin{figure}[t]
\includegraphics[width=0.5\textwidth]{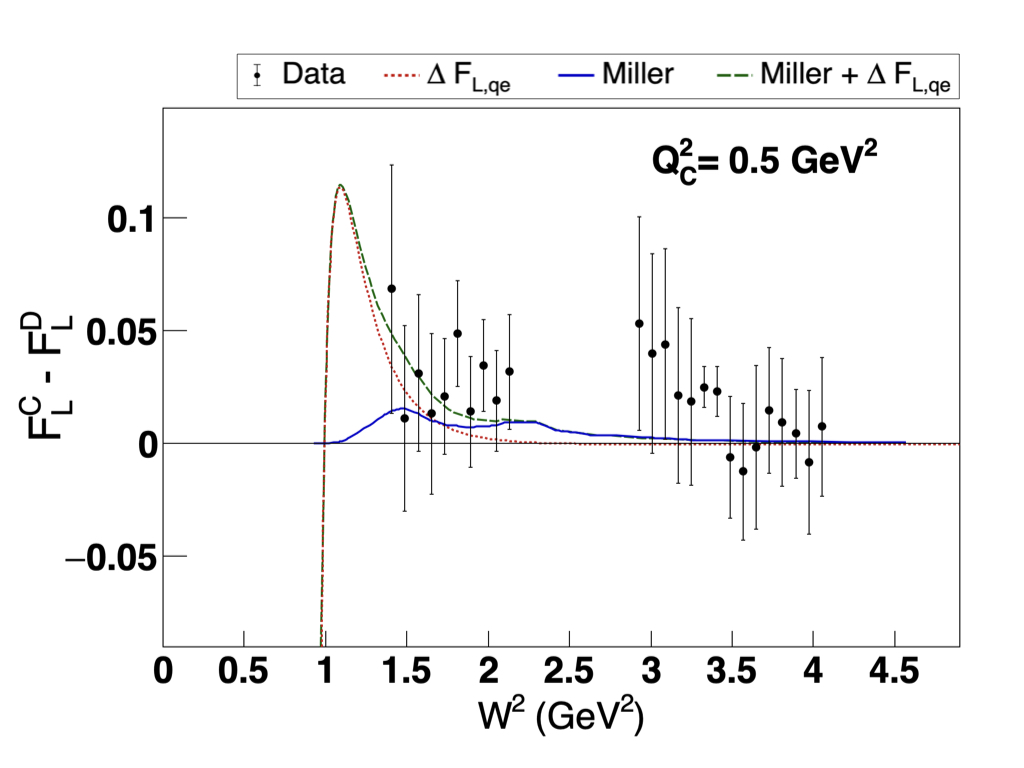}
\caption{The difference between the longitudinal structure functions for carbon and deuterium (black circles) compared to Gerald Miller's model \cite{Miller:2001yf} (blue line). The red line represents the difference between the quasielastic parts of the longitudinal structure functions as obtained from the universal fit \cite{Bodek:2022gli,Bodek:2023dsr}.}
\label{fig:Fig3}
\end{figure}
Values of $\Delta R >$ 0 could be  an indication of bosonic constituents (pions). 
In the model of Berger and  Coester \cite{Berger:1984na}, pions in nuclei are responsible for the ratio of the deep inelastic cross section on nuclei as compared to cross sections on free nucleons (especially a rise at {\it small} $x$) reported by the EMC Collaboration.  However, a subsequent publication by the Rochester-MIT-SLAC Collaboration \cite{Bodek:1983qn,Bodek:1983ec,Gomez:1993ri} showed that actually, there is a drop below unity in the ratio of $\sigma_A/\sigma_D$ at {\it small} $x$ (subsequently confirmed by the EMC Collaboration). In addition, meson clouds in nuclei would imply an enhancement of antiquarks in the nucleus which has been ruled out by Drell-Yan experiments on nuclear targets \cite{Alde:1990im}. \par

As a cross-check of the $R_C - R_D$ extraction presented above, $R$ is also extracted separately for deuterium  and $^{12}$C to verify the consistency and target dependence of $R$. For each target, $F_L$ and $F_1$  (and from these, $F_2$) are obtained from linear fits to the reduced cross section ($\sigma/\Gamma$) versus $\epsilon$ [Eq.~\ref{req}]. We require $\Delta \epsilon \equiv \epsilon_{\mathrm{max}} - \epsilon_{\mathrm{min}} \ge 0.25$; bins failing this leverage cut are omitted. Using $\epsilon'   = \epsilon /(1+\epsilon R_C)$, this is numerically comparable to $\Delta \epsilon' \ge 0.2$ over our kinematics. The uncertainty on $R$ includes the full covariance between $F_L$ and $F_1$ from the fits.

The individual cross section measurements have an average statistical uncertainty of $\sim$1.2\%. There is an overall systematic uncertainty in $R$ of $\pm 0.025$ arising from uncertainties in the theoretical formulation of the radiative corrections~\cite{Whitlow:1990gk}. Additional contribution include a $Q^2$ bin centering uncertainty (equal to 5\% of the applied correction) and a charge-symmetric background component, which contributes $\sim$ 3\% at large angles and high $W^2$. These sources combine to give a total point-to-point uncertainty in $\epsilon$ of $\sim$ 2.1\% on the cross sections used in the $R$ extraction. In addition, each target carries an overall normalization uncertainty of approximately 1.6\%, primarily from charge measurement, acceptance, and live-time corrections etc.

Figure.~\ref{Fig_R}  shows the extracted values of $R_D$ [Fig.~\ref{fig:R_d}] and $R_C$ [Fig.~\ref{fig:R_c}] versus $W^2$ at the eight values of $Q^2$ noted.   The resulting values of $R_C-R_D$ are in agreement with the primary analysis using Eq.~(\ref{req2}) (Fig.~\ref{fig:Fig_DR}).  Details of this analysis and investigation of individual structure functions ($F_1$ and $F_2$) for each target will be presented in a future communication.
\par

In a model by Gerald Miller~\cite{Miller:2001yf} pions in nuclei enhance the $F_L^A/F_L^D$ ratio at low $Q^2$ and {\it moderate and large} $x$. The model provided the values $F_L^C/F_L^D - 1$ in the deep-inelastic region, which we have converted converted to a difference, $F_L^C - F_L^D$, by multiplying the model by the inelastic value of $F_L^D$ obtained from the global fit \cite{Bodek:2022gli,Bodek:2023dsr}. The model predicts the largest effect to appear in a region near the peak of the $\Delta(1232)$, a region  which is obscured by the large contribution from quasielastic scattering  (Fig.~\ref{fig:Fig3}), which makes it difficult to test the model in this region.  At the higher values of $Q^2$ the  model is not able to describe our measurement of  $F_L^C - F_L^D$.

In summary, this is the first measurement of the nuclear dependence of $R$ in the resonance region, where no prior data exist. We find that $\Delta R >$ 0 at a level which is not explained by current theoretical models.

Acknowledgments--This material is based upon work supported by the U.S. Department of Energy, Office of Science, Office of Nuclear Physics under Contract No. 89243126CSC000213. This work was also supported by the U.S. Department of Energy under University of Rochester Grant No. DE-SC0008475 and by the U.S. National Science Foundation under Grant No. PHY-1914034. S. Alsalmi acknowledges support from the Ongoing Research Funding program (ORF-2026-1615), King Saud University, Riyadh, Saudi Arabia.

\bibliographystyle{apsrev4-1}

\bibliography{Jan5AB}

%

\end{document}

%% file: authors.tex
\author{S.~Alsalmi}
\affiliation{Department of Physics and Astronomy, King Saud University, Riyadh 11451, Saudi Arabia}

\author{I.~Albayrak}
\affiliation{Hampton University, Hampton, VA 23668, USA}

\author{A.~Ahmidouch}
\affiliation{North Carolina A\&T State University, Greensboro, NC 27411, USA}

\author{J.~Arrington}
\affiliation{Argonne National Laboratory, Argonne, IL 60439, USA}
\affiliation{Lawrence Berkeley National Laboratory, Berkeley, California 94720, USA}

\author{A.~Asaturyan}
\affiliation{A.I. Alikhanyan National Science Laboratory (Yerevan Physics Institute), Yerevan 0036, Armenia}
\affiliation{Thomas Jefferson National Accelerator Facility, Newport News, VA 23606, USA}

\author{A.~Bodek}
\affiliation{Department of Physics and Astronomy, University of Rochester, Rochester, NY 14627, USA}

\author{P.~Bosted}
\affiliation{Department of Physics, College of William \& Mary, Williamsburg, VA 23187, USA}

\author{R.~Bradford}
\affiliation{Argonne National Laboratory, Argonne, IL 60439, USA}
\affiliation{Department of Physics and Astronomy, University of Rochester, Rochester, NY 14627, USA}

\author{E.~Brash}
\affiliation{Christopher Newport University, Newport News, VA 23606, USA}

\author{A.~Bruell}
\affiliation{DFG, German Research Foundation, Kennedyallee 40, 53175 Bonn, Germany}

\author{C.~Butuceanu}
\affiliation{University of Regina, Regina, Saskatchewan S4S 0A2, Canada}

\author{M.~E.~Christy}
\affiliation{Hampton University, Hampton, VA 23668, USA}
\affiliation{Thomas Jefferson National Accelerator Facility, Newport News, VA 23606, USA}

\author{S.~J.~Coleman}
\affiliation{Department of Physics, College of William \& Mary, Williamsburg, VA 23187, USA}

\author{M.~Commisso}
\affiliation{University of Virginia, Charlottesville, VA 22904, USA}

\author{S.~H.~Connell}
\affiliation{University of Johannesburg, Auckland Park 2006, Johannesburg, South Africa}

\author{M.~M.~Dalton}
\affiliation{Thomas Jefferson National Accelerator Facility, Newport News, VA 23606, USA}

\author{S.~Danagoulian}
\affiliation{North Carolina A\&T State University, Greensboro, NC 27411, USA}

\author{A.~Daniel}
\affiliation{University of Houston, Houston, TX 77004, USA}

\author{D.~B.~Day}
\affiliation{University of Virginia, Charlottesville, VA 22904, USA}

\author{S.~Dhamija}
\affiliation{Florida International University, Miami, FL 33199, USA}

\author{J.~Dunne}
\affiliation{Mississippi State University, Mississippi State, MS 39762, USA}

\author{D.~Dutta}
\affiliation{Mississippi State University, Mississippi State, MS 39762, USA}

\author{R.~Ent}
\affiliation{Thomas Jefferson National Accelerator Facility, Newport News, VA 23606, USA}

\author{D.~Gaskell}
\affiliation{Thomas Jefferson National Accelerator Facility, Newport News, VA 23606, USA}

\author{A.~Gasparian}
\affiliation{North Carolina A\&T State University, Greensboro, NC 27411, USA}

\author{R.~Gran}
\affiliation{Department of Physics, University of Minnesota-Duluth, Duluth, MN 55812, USA}

\author{T.~Horn}
\affiliation{Thomas Jefferson National Accelerator Facility, Newport News, VA 23606, USA}

\author{Liting Huang}
\affiliation{Hampton University, Hampton, VA 23668, USA}

\author{G.~M.~Huber}
\affiliation{University of Regina, Regina, Saskatchewan S4S 0A2, Canada}

\author{C.~Jayalath}
\affiliation{Hampton University, Hampton, VA 23668, USA}

\author{M.~Johnson}
\affiliation{Argonne National Laboratory, Argonne, IL 60439, USA}
\affiliation{Northwestern University, Evanston, IL 60208, USA}

\author{M.~K.~Jones}
\affiliation{Thomas Jefferson National Accelerator Facility, Newport News, VA 23606, USA}

\author{N.~Kalantarians}
\affiliation{Virginia Union University, Richmond, VA 23220, USA}

\author{A.~Liyanage}
\affiliation{Hampton University, Hampton, VA 23668, USA}

\author{C.~E.~Keppel}
\affiliation{Hampton University, Hampton, VA 23668, USA}

\author{E.~Kinney}
\affiliation{University of Colorado, Boulder, CO, USA}

\author{Y.~Li}
\affiliation{Hampton University, Hampton, VA 23668, USA}

\author{S.~Malace}
\affiliation{Duke University, Dept. of Physics, Box 90305, Durham, NC 27708}

\author{V.~Mamyan}
\affiliation{University of Virginia, Charlottesville, VA 22904, USA}

\author{S.~Manly}
\affiliation{Department of Physics and Astronomy, University of Rochester, Rochester, NY 14627, USA}

\author{P.~Markowitz}
\affiliation{Florida International University, Miami, FL 33199, USA}

\author{J.~Maxwell}
\affiliation{University of Virginia, Charlottesville, VA 22904, USA}

\author{N.~N.~Mbianda}
\affiliation{University of Johannesburg, Auckland Park 2006, Johannesburg, South Africa}

\author{K.~S.~McFarland}
\affiliation{Department of Physics and Astronomy, University of Rochester, Rochester, NY 14627, USA}

\author{M.~Meziane}
\affiliation{Department of Physics, College of William \& Mary, Williamsburg, VA 23187, USA}

\author{Z.~E.~Meziani}
\affiliation{Department of Physics, Temple University, Philadelphia, PA 19122, USA}

\author{G.~B.~Mills}
\affiliation{Los Alamos National Laboratory, Los Alamos, NM 87545, USA}

\author{H.~Mkrtchyan}
\affiliation{A.I. Alikhanyan National Science Laboratory (Yerevan Physics Institute), Yerevan 0036, Armenia}

\author{A.~Mkrtchyan}
\affiliation{A.I. Alikhanyan National Science Laboratory (Yerevan Physics Institute), Yerevan 0036, Armenia}

\author{J.~Mulholland}
\affiliation{University of Virginia, Charlottesville, VA 22904, USA}

\author{J.~K.~Nelson}
\affiliation{Department of Physics, College of William \& Mary, Williamsburg, VA 23187, USA}

\author{G.~Niculescu}
\affiliation{James Madison University, Harrisonburg, VA 22801, USA}

\author{I.~Niculescu}
\affiliation{James Madison University, Harrisonburg, VA 22801, USA}

\author{L.~Pentchev}
\affiliation{Department of Physics, College of William \& Mary, Williamsburg, VA 23187, USA}

\author{A.~Puckett}
\affiliation{Massachusetts Institute of Technology, Cambridge, MA 02139, USA}
\affiliation{Los Alamos National Laboratory, Los Alamos, NM 87545, USA}

\author{V.~Punjabi}
\affiliation{Norfolk State University, Norfolk, VA 23504, USA}

\author{I.~A.~Qattan}
\affiliation{Khalifa University of Science and Technology, Abu Dhabi 127788, UAE}

\author{P.~E.~Reimer}
\affiliation{Argonne National Laboratory, Argonne, IL 60439, USA}

\author{J.~Reinhold}
\affiliation{Florida International University, Miami, FL 33199, USA}

\author{V.~M.~Rodriguez}
\affiliation{University of Houston, Houston, TX 77004, USA}

\author{O.~Rondon-Aramayo}
\affiliation{University of Virginia, Charlottesville, VA 22904, USA}

\author{M.~Sakuda}
\affiliation{High Energy Accelerator Research Organization (KEK), Tsukuba, Ibaraki 305-0801, Japan}

\author{W.~K.~Sakumoto}
\affiliation{Department of Physics and Astronomy, University of Rochester, Rochester, NY 14627, USA}

\author{E.~Segbefia}
\affiliation{Hampton University, Hampton, VA 23668, USA}

\author{T.~Seva}
\affiliation{University of Zagreb, Zagreb, Croatia}

\author{I.~Sick}
\affiliation{University of Basel, CH-4056 Basel, Switzerland}

\author{K.~Slifer}
\affiliation{University of New Hampshire, Durham, NH 03824, USA}

\author{G.~R.~Smith}
\affiliation{Thomas Jefferson National Accelerator Facility, Newport News, VA 23606, USA}

\author{J.~Steinman}
\affiliation{Department of Physics and Astronomy, University of Rochester, Rochester, NY 14627, USA}

\author{P.~Solvignon}
\affiliation{Argonne National Laboratory, Argonne, IL 60439, USA}

\author{V.~Tadevosyan}
\affiliation{A.I. Alikhanyan National Science Laboratory (Yerevan Physics Institute), Yerevan 0036, Armenia}

\author{S.~Tajima}
\affiliation{University of Virginia, Charlottesville, VA 22904, USA}

\author{V.~Tvaskis}
\affiliation{University of Winnipeg, Winnipeg, Manitoba R3B 2E9, Canada}

\author{W.~F.~Vulcan}
\affiliation{Thomas Jefferson National Accelerator Facility, Newport News, VA 23606, USA}

\author{T.~Walton}
\affiliation{Hampton University, Hampton, VA 23668, USA}

\author{F.~R.~Wesselmann}
\affiliation{Norfolk State University, Norfolk, VA 23504, USA}

\author{S.~A.~Wood}
\affiliation{Thomas Jefferson National Accelerator Facility, Newport News, VA 23606, USA}

\author{Zhihong~Ye}
\affiliation{Hampton University, Hampton, VA 23668, USA}

\collaboration{The JUPITER Collaboration: JLab E02-109 E04-001 E06-009}
\noaffiliation